%% file: GADT-story.tex
\UseRawInputEncoding
\documentclass[submission,copyright,creativecommons]{eptcs}
\providecommand{\event}{LSFA 2021}
\usepackage{breakurl}             
\usepackage{underscore}           

\usepackage[utf8]{inputenc}
\usepackage{ccicons}
\usepackage{verbatim}

\usepackage{amsmath}
\usepackage{amsthm}
\usepackage{amscd}
\usepackage{xcolor}

\usepackage{bbold}
\usepackage{url}
\usepackage{upgreek}

\usepackage{lipsum}
\usepackage{tikz-cd}
\usetikzlibrary{cd}
\usetikzlibrary{calc}
\usetikzlibrary{arrows}

\usepackage{bussproofs}
\EnableBpAbbreviations

\input{macros}

\input{mytheorem}

\newtheorem{example}[thm]{Example}

\newtheorem{definition}[thm]{Definition}

\renewcommand{\greyout}[1]{}

\renewcommand{\id}{\mathit{id}}

\title{GADTs, Functoriality,  Parametricity: Pick Two}
\author{Patricia Johann, Enrico Ghiorzi, and Daniel Jeffries
\institute{Appalachian State University}
\email{johannp@appstate.edu, ghiorzie@appstate.edu,
  jeffriesd@appstate.edu}}

\def\titlerunning{GADTs, Functoriality,  Parametricity: Pick Two}
\def\authorrunning{P. Johann, E. Ghiorzi, and D. Jeffries}

\begin{document}
\maketitle

\begin{abstract}
GADTs can be represented either as their Church encodings {\em \`a la}
Atkey, or as fixpoints {\em \`a la} Johann and Polonsky. While a GADT
represented as its Church encoding need not support a $\mathsf{map}$
function satisfying the functor laws, the fixpoint representation of a
GADT must support such a $\mathsf{map}$ function even to be
well-defined. The two representations of a GADT thus need not be the
same in general. This observation forces a choice of representation of
data types in languages supporting GADTs.  In this paper we show that
choosing whether to represent data types as their Church encodings or
as fixpoints determines whether or not a language supporting GADTs can
have parametric models. This choice thus has important consequences
for how we can program with, and reason about, these advanced data
types.
\end{abstract}

\maketitle

\section{Introduction}\label{sec:intro}

There are two standard ways to represent data types when studying
modern functional languages: {\em i}) as their Church encodings in a
(possibly higher-kinded) extension of System F~\cite{gir72}, the
calculus at the core of all such languages, and {\em ii}) by
augmenting System F with primitives for constructing them directly as
fixpoints.  Models in which the Church encoding and the fixpoint
representation of every algebraic data type (ADT) are semantically
equivalent include the operationally-based model of~\cite{pit98,pit00}
and the categorical models of~\cite{joh02,joh03,jgj21}.  In the
categorical model of~\cite{jgj21} this semantic equivalence also holds
for the syntactic generalization of ADTs known as nested
types~\cite{bm98}. Generalized algebraic data types
(GADTs)~\cite{pvww06} generalize nested types --- and thus further
generalize ADTs --- syntactically (see Section~\ref{sec:gadts} below):

\vspace*{-0.1in}

\[\begin{tikzcd}[column sep = huge]
\text{\fbox{$\mathsf{ADTs}$}}\;
\ar[r,hookrightarrow,"\text{syntactically}","\text{generalized by}"']
& \;\text{\fbox{$\mathsf{nested~types}$}}\; \ar[r,hookrightarrow,"\text{syntactically}","\text{generalized by}"']
& \;\text{\fbox{$\mathsf{GADTs}$}}
\end{tikzcd}\]

\vspace*{-0.05in}

\noindent
This sequence of syntactic generalizations suggests that a
corresponding sequence of semantic generalizations might also hold.
However, this is not the case. Specifically, the semantic equivalence
between Church encodings and fixpoint representations that holds for
ADTs and nested types need not hold for GADTs. In a language
supporting GADTs we must therefore choose whether to represent data
types as their Church encodings or as fixpoints.  The main result of
this paper shows that the choice of representation of data types in a
language supporting GADTs determines whether or not that language can
have parametric models~\cite{rey83}. It thus determines whether or not
GADTs can enjoy consequences of parametricity such as representation
independence~\cite{adr09,dnb12}, equivalences between
programs~\cite{hd11}, deep induction principles~\cite{jg22,jp20},
and useful (``free'') theorems about programs derived from their types
alone~\cite{wad89}.  This result is unintuitive, novel, and
surprising.

Consider the GADT $\mathsf{Seq}$ defined by\footnote{Although our
  development applies to GADTs in any language, we will use Haskell
  syntax for the code in this paper.}
\begin{equation}\label{eq:seq}
\begin{array}{l}
\mathsf{data\, Seq\,a\,where}\\
\mathsf{\;\;\;\;\;\;\;\;Const ::\, a \to Seq\,a}\\
\mathsf{\;\;\;\;\;\;\;\;Pair\,\,\,\,\, ::\, Seq \,a \to Seq\,b \to
  Seq\,(a \times b)}\\ 
\end{array}
\end{equation}
\noindent
This GADT comprises sequences of any type $\mathsf{a}$ and sequences
obtained by pairing the data in two already existing sequences. When
represented as its Church encoding, this GADT contains no data
elements other than these. More generally, a GADT's Church encoding
contains exactly those data elements that are representable by its
syntax. By contrast, the fixpoint representation of a GADT sees that
data type as the {\em functorial completion}~\cite{jp19} of its
syntax.  Completing a GADT's syntax from a funct{\em ion} on types to
a funct{\em or} on types is necessary if its interpretation is to be a
fixpoint~\cite{tfca}, so functoriality is inherent in fixpoint
representations of data types. Since being a functor entails
supporting a map function satisfying the functor laws, the fixpoint
representation of a data type must include the entire ``map closure''
of its syntax.  For example, the fixpoint representation of
$\mathsf{Seq}$ contains not just those data elements representable by
its syntax, but also all data elements of the form
$\mathsf{map\,f\,s}$ for all types $\mathsf{t_1}$ and $\mathsf{t_2}$,
all functions $\mathsf{f :: t_1 \to t_2}$ definable in the language and
all $\mathsf{s :: Seq\,t_1}$, as well as all data elements of the form
$\mathsf{map\,g\,s'}$ for each appropriately typed function
$\mathsf{g}$ and each element $\mathsf{s'}$ already added to the data
type, and so on. Functorial completion for $\mathsf{Seq}$ adds, in
particular, data elements of the form
$\mathsf{map\,g\,(Pair\,u_1\,u_2)}$ even though these may not
themselves be of the form $\mathsf{Pair\,v_1\,v_2}$ for any terms
$\mathsf{v_1}$ and $\mathsf{v_2}$. Importantly, functorial completion
adds no new data elements to the syntax of a GADT that is an ADT or
other nested type, which perfectly explains why Church encodings and
fixpoint representations coincide for ADTs and other nested
types. However, the two representations are not the same for GADTs
that are not nested types.\looseness=-1

Even so, does it really matter how we represent GADTs?

A GADT programmer is likely to use GADTs precisely {\em because} they
exhibit different behaviors at different types, and thus to consider a
GADT to be completely specified by its syntax. When used this way, the
shape of a particular element of a GADT is actually {\em determined}
by the data it contains. As a result, it cannot be expected to support
a functorial $\mathsf{map}$ function like ADTs and nested
types. Indeed, the definition for $\mathsf{Seq}$ above specifies that
an element of the form $\mathsf{Pair\,u_1\,u_2}$ must have the shape
of a sequence of data of pair type rather than a sequence of data of
arbitrary type $\mathsf{e}$.  The clause of $\mathsf{map}$ for the
$\mathsf{Pair}$ constructor should therefore feed $\mathsf{map}$ a
function $\mathsf{f :: (a \times b) \to e}$ and a term of the form
$\mathsf{Pair \,u_1\,u_2}$ for $\mathsf{u_1 :: Seq\,a}$ and
$\mathsf{u_2 :: Seq\,b}$, and produce a term $\mathsf{Pair\,v_1\,v_2}$
for some appropriately typed terms $\mathsf{v_1}$ and
$\mathsf{v_2}$. However, it is not clear how to achieve this since
$\mathsf{e}$ need not necessarily be a product type. And even if
$\mathsf{e}$ {\em were} known to be of the form $\mathsf{w \times z}$,
we still wouldn't necessarily have a way to produce data of type
$\mathsf{w \times z}$ from only $\mathsf{f :: a \times b \to w \times
  z}$ and $\mathsf{u_1}$ and $\mathsf{u_2}$ unless we knew, e.g., that
$\mathsf{f}$ was a pair of functions $\mathsf{(f_1 :: a \to w, f_2 ::
  b \to z)}$. When it is intended to capture this kind of
non-uniformity, a GADT cannot be regarded as a data-independent
container that can be filled with data of any type in the same way
that ADTs and nested types can. In this situation, a GADT cannot
denote a functor, and so must necessarily be represented as its
Church encoding.

A semanticist, on the other hand, is likely to expect GADTs to
generalize ADTs semantically --- i.e., to have the same kind of
functorial semantics that ADTs and nested types
have~\cite{bm98,jg07,jp19} --- and thus to be useable as a container
filled with data that can be changed without changing the shape of the
container itself.  For a GADT to be used in this way, its syntax must
reflect functoriality. Since the Church encoding of a GADT that is not
a nested type does not denote a functor, to have a functorial
semantics such a GADT must be viewed as its fixpoint
representation. The functorial completion inherent in the fixpoint
representation of a GADT adds to its syntax those, and only those,
data elements needed to support a functorial $\mathsf{map}$ function.
For example, let $\mathsf{1}$ be the unit type whose single element is
also denoted by $\mathsf{1}$, and let $\mathsf{G}$ be the GADT defined
by\looseness=-1
\begin{equation}\label{eq:G}
\begin{array}{l}
\mathsf{data\,G\,a\,where}\\
\mathsf{\;\;\;\;\;\;\;\;C :: G\,1}
\end{array}
\end{equation}
Then the functorial completion of $\mathsf{G}$ includes elements at
any instance $\mathsf{G\,a}$ for any type $\mathsf{a}$ that is not the
empty type $\mathsf{0}$ (since there is always a function from
$\mathsf{1}$ to such an $\mathsf{a}$), but includes no elements at
instance $\mathsf{G\,0}$. Indeed, $\mathsf{G\,0}$ is not inhabited via
the syntactic specification of $\mathsf{G}$, and is not inhabited via
functorial completion because it is not possible to define a function
from $\mathsf{1}$ into $\mathsf{0}$.

The key observation of this paper is that, while the viewpoints of the
GADT programmer and the semanticist are both valid, the two are
irreconcilable. Importantly, which of our two representations of data
types is adopted in any particular setting has significant
consequences for the ways GADTs can be used and reasoned about
there. In particular, the way that GADTs are represented has deep
implications for parametric reasoning about them. Specifically, a
programmer who views GADTs as their Church encodings cannot safely use
program transformations or reasoning principles that involve
$\mathsf{map}$ functions for them, although they may be able to
program with and reason about GADTs using other consequences of
parametricity, such as type inhabitation results. On the other hand, a
semanticist who views GADTs as fixpoints will have all
naturality-based program transformations and reasoning principles for
GADTs at their disposal since these all derive from functoriality. But
since, as we show in Section~\ref{sec:par}, no parametric model can be
constructed for fixpoint representations of GADTs, non-naturality
consequences of parametricity will not necessarily hold for them.
Overall, we show that, as with software engineering's iron
triangle~\cite{iron-triangle}, we can have any two of {\em GADTs},
{\em functoriality}, and {\em parametricity} we like, but we cannot
have all three.

The goal of this paper is to show how the above observations can be
made precise, and thereby to answer the question we posed above in the
affirmative: Yes, it really does matter how we represent GADTs.

\section{Representations of Algebraic Data Types}\label{sec:adts}

A (polynomial) {\em algebraic data type} (ADT) has the form
\[\mathsf{T\,a} = \mathsf{C_1 t_{11}}...\mathsf{t_{1k_1}}\, |\, ...\, |\,
\mathsf{C_n t_{n1}}...\mathsf{t_{nk_n}}\] where each $\mathsf{t_{ij}}$
is a type depending only on $\mathsf{a}$. Such a data type can be
thought of as a ``container'' for data of type $\mathsf{a}$. The data
in an ADT are arranged at various {\em positions} in its underlying
{\em shape}, which is determined by the types of its {\em
  constructors} $\mathsf{C_1},...,\mathsf{C_n}$. An ADT's constructors
are used to build the data values of the data type, as well as to
analyze those values using {\em pattern matching}. ADTs are used
extensively in functional programming to structure computations, to
express invariants of the data over which computations are defined,
and to ensure the type safety of programs specifying those
computations.

List types are the quintessential examples of ADTs. The shape of the
container underlying the
type
\[\mathsf{List\,a} = \mathsf{Nil} \,|\, \mathsf{Cons\,a\,(List\,a)}\]
is determined by the types of its two constructors $\mathsf{Nil\, ::\,
  List\, a}$ and $\mathsf{Cons \,::\,a \to \List\,a\to
  List\,a}$. These constructors specify that the data in a list of
type $\mathsf{List\,a}$ are arranged linearly. The shape underlying
the type $\mathsf{List\,a}$ is therefore given by the set $\nat$ of
natural numbers, with each natural number representing a choice of
length for a list structure, and the positions in a structure of shape
$\mathsf{n}$ are given by natural numbers ranging from $\mathsf{0}$ to
$\mathsf{n-1}$. Since the type argument to every occurrence of the
type constructor $\mathsf{List}$ in the right-hand side of the above
definition is the same as the type instance being defined on its
left-hand side, the type $\mathsf{List\,a}$ enforces the invariant
that all of the data in a structure of this type have the same type
$\mathsf{a}$. In a similar way, the tree type
\[\mathsf{Tree\, a} = \mathsf{Leaf\,a}
\,|\,\mathsf{Node\,(Tree\,a)\,a\,(Tree\,a)}\] of binary trees has as
its underlying shape the type of binary trees of units, and the
positions in a structure of this type are given by sequences of L (for
``left'') and R (for ``right'') navigating a path through the
structure. The type $\mathsf{Tree\, a}$ enforces the invariant that
all of the data at the nodes and leaves in a structure of this type
have the same type $\mathsf{a}$.

Since the shape of an ADT structure --- i.e., a structure whose type
is an instance of an ADT --- is independent of the type of data it
contains, ADTs can be defined polymorphically. As a result, an ADT
structure containing data of type $\mathsf{a}$ can be transformed into
another ADT structure of the exact same shape containing data of
another type $\mathsf{b}$ simply by applying a given function
$\mathsf{f \,::\,a \to b}$ to each of its elements. Moreover, every ADT
$\mathsf{T}$ can be made an instance of Haskell's $\mathsf{Functor}$
class by defining a type-and-data-uniform, structure-preserving,
data-changing function $\mathsf{map_T}$ for it.\footnote{We write
  $\mathsf{map_T}$, or simply $\mathsf{map}$ when $\mathsf{T}$ is
  clear from context, for the function $\mathsf{fmap :: (a \to b) \to
    (T\,a \to T\,b)}$ witnessing that a type constructor $\mathsf{T}$
  is an instance of Haskell's $\mathsf{Functor}$ class. We
  emphasize that $\mathsf{fmap}$ functions in Haskell are intended to
  satisfy syntactic reflections of the functor laws ---
  i.e., preservation of identity functions and composition of
  functions --- even though this is not enforced by the compiler and
  is instead left to the good intentions of the programmer.}
Then, given a type-independent way of rearranging an ADT
structure's shape $\mathsf{T\,a}$ into the shape for another ADT
structure $\mathsf{T'\,a}$, we get the same structure of type
$\mathsf{T'\,b}$ regardless of whether we first rearrange the original
structure of type $\mathsf{T\,a}$ into one of type $\mathsf{T'\,a}$
and then use $\mathsf{map_{T'}}$ to convert that resulting structure
to one of type $\mathsf{T'\,b}$, or we first use $\mathsf{map_T}$ to
convert the original structure of type $\mathsf{T\,a}$ to one of type
$\mathsf{T\,b}$ and then rearrange that resulting structure into one
of type $\mathsf{T'\,b}$. For example, if $\mathsf{f :: a \to b}$,
$\mathsf{t \,::\,Tree\,a}$, and $\mathsf{g :: Tree\,a\to List\,a}$
arranges trees into lists in a type-independent way, then we have the
following rearrange-transform property:
\[\mathsf{map_{List}\,f\,(g \, t) \,=\, g\,(map_{Tree}\,f\,t)}\]

\subsection{Church Encodings of ADTs}

One way to represent ADTs is as their Church encodings. A Church
encoding is a representation of a data type as a function in a pure
lambda calculus, such as System F and its higher-kinded
extensions. They, together with other related encodings, have recently
been popularized as various {\em visitor patterns} in object-oriented
programming~\cite{gon21,owg08}.

Church encodings of ADTs can be defined in any language that supports
functions. They can therefore be used to represent ADTs in languages
that do not support primitives for sum types, product types, or
recursion. The Church encodings of the ADTs $\mathsf{List\,a}$ and
$\mathsf{Tree\,a}$, for example, are
\[\mathsf{List\,a} = \mathsf{\forall b.\,b \to (a \to b \to b) \to b}\]
and 
\[\mathsf{Tree\,a} = \mathsf{\forall b.\,(a \to b) \to (b \to a \to b
  \to b) \to b}\] respectively. The argument types in a Church
encoding of an ADT are abstractions of the types of the ADT's
constructors. For instance, $\mathsf{b}$ abstracts the type
$\mathsf{List\,a}$ of the constructor $\mathsf{Nil}$ for lists, and
$\mathsf{a \to b \to b}$ abstracts the type $\mathsf{a \to List\,a \to
  List\,a}$ of the constructor $\mathsf{Cons}$.

Because the types of the ``abstract constructors'' for an ADT are
uniform in their argument types, it is always possible to Church
encode the type constructors themselves as well. For example, the
Church encodings of the type constructors $\mathsf{List}$ and
$\mathsf{Tree}$ are
\[\mathsf{List} = \mathsf{\forall a. \forall b.\,b \to (a \to b \to b)
  \to b}\] 
and 
\[\mathsf{Tree} = \mathsf{\forall a. \forall b.\,(a \to b) \to (b \to
  a \to b \to b) \to b}\] respectively. This fact is what allows an
ADT's associated type constructor to be made an instance of Haskell's
$\mathsf{Functor}$ class. For example, the $\mathsf{map}$ function
from Haskell's standard library makes the type constructor
$\mathsf{List}$ an instance of the $\mathsf{Functor}$ class, and for
$\mathsf{Tree}$ we can define
\[\begin{array}{lll}
\mathsf{map_{Tree}} & \mathsf{::} & \mathsf{(a \to b) \to Tree\,a \to Tree \,b}\\
\mathsf{map_{Tree}\, f\, (Leaf\,x)} & \mathsf{=} & \mathsf{Leaf\,(f\,x)}\\
\mathsf{map_{Tree}\, f\, (Node\,t_1\,x\,t_2)} & \mathsf{=} &
\mathsf{Node\,(map_{Tree}\, f \,t_1)\,(f\,x)\,(map_{Tree}\,f\,t_2)}\\
\end{array}\]
Note that the Church encoding of an ADT carries with it no expectation
whatsoever that such a type-and-data-uniform, structure-preserving,
data-changing $\mathsf{map}$ function can be defined or that, if one
can, it will satisfy a rearrange-transform property. (Of course, such
a $\mathsf{map}$ function can always be defined for any ADT precisely
because its Church encoding and its fixpoint representation coincide,
and this $\mathsf{map}$ function will necessarily satisfy a
rearrange-transform property.)


\subsection{ADTs as Fixpoints} 

By contrast, the ability to define such a $\mathsf{map}$ function is
inherent in the view of ADTs as fixpoints. Such a view is possible in
any language that supports primitives for sum types, product types,
and recursion. In such a language, the fixpoint representations of
  the ADTs $\mathsf{List\,a}$ and $\mathsf{Tree\,a}$ are
\begin{equation}\label{eq:list}
  \mathsf{List\,a} = \mathsf{\mu X.\, 1 + a \times X}
\end{equation}
and 
\begin{equation}\label{eq:tree}
  \mathsf{Tree\,a} = \mathsf{\mu X.\, a + X \times a \times X}
\end{equation}
respectively, where $\mathsf{\mu}$ is a primitive fixpoint operator.

Fixpoint representations capture in syntax the fact that ADTs can be
considered as fixpoints of functors. For example, $\mathsf{List\,a}$ is
indeed a fixpoint of $\mathsf{F_{List\,a}}$, where
$\mathsf{F_{List\,a}\,X} = \mathsf{1 + a \times X}$
by~\eqref{eq:list}. That is,
\[\mathsf{List\,a} = \mathsf{1 + a \times List\,a}\] 
since every element of $\mathsf{List\,a}$ is either empty or is
obtained by consing an element of type $\mathsf{a}$ onto an
already-existing structure of type $\mathsf{List\,a}$. In fact, this
fixpoint equation is just a rewriting of the Haskell data type
declaration for $\mathsf{List\,a}$. We therefore have that
$\mathsf{List\, a} = \mathsf{\mu F_{List\,a}}$ is modeled by $\mu
F_{\mathit{List}\,a}$, where the functor $F_{\mathit{List}\,a}$ models
the type constructor $\mathsf{F_{List\,a}}$.\footnote{Throughout this
  paper, we use $\mathtt{sans\; serif}$ font for program text and
  $\mathit{math \; italic}$ font for semantic objects.}  Similarly,
the ADT $\mathsf{Tree\,a}$ can be seen to be a fixpoint of
$\mathsf{F_{Tree\,a}}$, where $\mathsf{F_{Tree\,a}\,X} = \mathsf{a + X
  \times a \times X}$ by~\eqref{eq:tree}, so that $\mathsf{Tree\, a} =
\mathsf{ \mu F_{Tree\,a}}$ is modeled by $\mu F_{\mathit{Tree}\,a}$ if
$F_{\mathit{Tree}\,a} $ models $\mathsf{F_{Tree\,a}}$.
 
These kinds of fixpoint equations are entirely sensible at the level
of types. But to ensure that the syntactic fixpoint representing an
ADT actually denotes a semantic object computed as a semantic
fixpoint, the semantic fixpoint calculation must converge.  If, as is
typical, we interpret our types as sets, then the fixpoint being taken
must be of a funct{\em or} on the category $\mathit{Set}$ of sets and
functions between them, rather than of a mere funct{\em ion} between
sets~\cite{tfca}.  That is, the function $F$ interpreting the type
constructor $\mathsf{F}$ constructing the body of a syntactic fixpoint
must be a functor, i.e., must not only have an action on sets, but
must also have a {\em functorial action} on functions between
sets. Reflecting this requirement back into syntax gives that
$\mathsf{F}$ must support a
$\mathsf{map}$ function satisfying the functor laws. That is,
$\mathsf{F}$ must be an instance of Haskell's $\mathsf{Functor}$ class
(with the aforementioned caveat about the functor laws).

Requiring $F$ to be a functor ensures that the interpretation $\mu F$
of the ADT $\mathsf{T\,a} = \mathsf{\mu F}$ exists. But to ensure that
$\mu F$ is itself a functor, so that the type constructor $\mathsf{T}$
associated with $\mathsf{T\,a}$ also supports its own $\mathsf{map}$
function, we can require that $F$ be a functor on the category
$\mathit{Set}^\mathit{Set}$ of functors and natural transformations on
$\mathit{Set}$. That is, $F$ must be a {\em higher-order} functor on
$\mathit{Set}$. Writing $H$ in place of $F$ to emphasize that it is
higher-order, and reflecting this requirement back into syntax, we
have that $\mathsf{T\,a} = \mathsf{(\mu H)\,a}$ for the ``type
constructor constructor'' $\mathsf{H}$ that supports
suitable\footnote{The $\mathsf{map}$ function for $\mathsf{H}$ is
  intended to satisfy syntactic reflections of the functor laws in
  $\mathit{Set}^\mathit{Set}$ --- i.e., preservation of identity
  natural transformations and composition of natural transformations
  --- and the $\mathsf{map}$ function for $\mathsf{H\,F}$ is intended
  to satisfy syntactic reflections of the functor laws in
  $\mathit{Set}$, even though there is no mechanism in Haskell for
  enforcing this.} $\mathsf{map}$ functions.

A concrete example is given by the ADT $\mathsf{List\,a}$. This type
is modeled as the fixpoint $\mu F_{\mathit{List\,a}}$ of the
first-order functor whose action on sets is given by
$F_{\mathit{List}\,a}\,X = 1 + a \times X$ and whose action on
functions is given by $F_{\mathit{List}\,a}\,f = 1 + \id_a \times
f$. The type constructor $\mathsf{List}$ is modeled by the functor
that is the fixpoint $\mu H$ of the higher-order functor $H$ whose
action on a functor $F$ is given by the functor $H\,F$ whose actions
on sets and functions between them are given by $H\,F\,X = 1 + X
\times F\,X$ and $H\,F\,f = \id_1 + f \times F\,f$, respectively, and
whose action on a natural transformation $\eta$ is the natural
transformation whose component at $X$ is given by $(H\,\eta)_X = \id_1
+ \id_X \times \eta_X$. Reflecting the functorial action of $\mu H$
back into syntax gives exactly Haskell's built-in $\mathsf{map}$
function as the type-and-data-uniform, structure-preserving,
data-changing function associated with $\mathsf{List}$.

\vspace*{0.05in}

In any parametric model, the Church encoding and the fixpoint
representation of an ADT or a nested type will necessarily be
semantically equivalent. But whereas it is impossible even to {\em
  state} the rearrange-transform property for their Church encodings
(unless functorial $\mathsf{map}$ functions have somehow been defined
for them), for their fixpoint representations such a property is
simply the reflection back into syntax of the instance of naturality
for the type-independent function that rearranges structures of type
$\mathsf{T\,a}$ into ones of type $\mathsf{T'\,a}$ and the
structure-preserving, data-changing functions $\mathsf{map_T\,f}$ and
$\mathsf{map_{T'}\,f}$ for a function $\mathsf{f :: a \to b}$, where
$\mathsf{T}$ and $\mathsf{T'}$ are the type constructors associated
with these ADTs, respectively.

\section{Representations of GADTs}\label{sec:gadts}

{\em Generalized algebraic data types} (GADTs)~\cite{pvww06} relax the
restriction on the type instances appearing in a data type
definition. The special form of GADTs known as {\em nested
  types}~\cite{bm98} allow the data constructors of a GADT to take as
arguments data whose types involve type instances of the GADT other
than the one being defined. However, the return type of each
constructor of a nested type must still be precisely the one being
defined. This is illlustrated by the definition
\[\mathsf{PTree \, a = PLeaf \,a \,|\, PNode \,(PTree\,(a \times
  a))}\] of the nested type $\mathsf{PTree\,a}$ of perfect trees,
which introduces the data constructors $\mathsf{PLeaf :: a \to
  PTree\,a}$ and $\mathsf{PNode :: PTree \,(a\times a) \to
  PTree\,a}$. It enforces not only the invariant that all of the data
in a structure of type $\mathsf{PTree\,a}$ is of the same type
$\mathsf{a}$, but also the invariant that all perfect trees have
lengths that are powers of 2. GADTs that are not nested types allow
their constructors both to take as arguments {\em and return as
  results} data whose types involve type instances of the GADT other
than the one being defined. An example is the GADT $\mathsf{Seq}$
given in~\eqref{eq:seq}. Since the return type of the data constructor
$\mathsf{Pair}$ is not of the form $\mathsf{Seq\,a}$ for any variable
$\mathsf{a}$, $\mathsf{Seq}$ is a GADT that is not a nested type.

By contrast with the ADT $\mathsf{List\, a}$, where the type parameter
$\mathsf{a}$ is integral to the type being defined, the type parameter
$\mathsf{a}$ appears in both $\mathsf{PTree\,a}$ and $\mathsf{Seq\,a}$
as a ``dummy'' parameter used only to give the kind $\mathsf{* \to *}$ of
the type constructors $\mathsf{PTree}$ and $\mathsf{Seq}$. This is
explicitly captured in the alternative ``kind signature'' Haskell
syntax, which represents $\mathsf{PTree}$ and $\mathsf{Seq}$ as
\[\begin{array}{l}
\mathsf{data\, PTree :: * \to *\,where}\\
\mathsf{\;\;\;\;\;\;\;\;PLeaf \,\,\,\;::\, a \to PTree\,a}\\
\mathsf{\;\;\;\;\;\;\;\;PNode\,\, ::\, PTree \,(a \times a) \to
  PTree\,a}\\ 
\end{array}\]
\noindent
and
\[\begin{array}{l}
\mathsf{data\, Seq :: * \to *\,where}\\
\mathsf{\;\;\;\;\;\;\;\;Const ::\, a \to Seq\,a}\\
\mathsf{\;\;\;\;\;\;\;\;Pair\,\,\,\,\; ::\, Seq \,a \to Seq\,b \to
  Seq\,(a \times b)}\\ 
\end{array}\]
respectively. A GADT --- even a nested type --- thus does not define a {\em family
  of inductive types}, one for each type argument, like an ADT does,
but instead defines an entire family of types that must be constructed
simultaneously. That is, a GADT defines an {\em inductive family of
  types}.

GADTs that are not nested types are used in precisely those situations
in which different behaviors at different instances of a data type are
desired. This is achieved by allowing the programmer to give the type
signatures of the GADT's data constructors independently --- as is
made explicit by the alternative syntax above --- and then using
pattern matching to force the desired type refinement. Applications of
GADTs include generic programming, modeling programming languages via
higher-order abstract syntax, maintaining invariants in data
structures, and expressing constraints in embedded domain-specific
languages. GADTs have also been used, e.g., to implement tagless
interpreters~\cite{pl04,pr06,pvww06}, to improve memory
performance~\cite{min15}, and to design APIs~\cite{pen20}.

\subsection{Church Encodings of GADTs}\label{sec:CEs}

The syntax of GADTs allows non-variable type arguments in the return
types of their data constructors. This establishes a strong connection
between a GADT's shape and the data it contains. With ADTs, we first
choose the shape of the container and then fill that container with
data of whatever type we like; critically, the choice of shape is
independent of the data to be stored. With GADTs, however, the shape
of the container may actually {\em depend} on (the type of) the data
to be contained. For example, $\mathsf{Const}$ can create data of any
shape $\mathsf{Seq\,a}$, but $\mathsf{Pair}$ can produce data of shape
$\mathsf{Seq\,a}$ only if $\mathsf{a}$ is a pair type. As a result,
modifying the data in a GADT's container may change the shape of that
container, or even produce an ill-typed result.

To determine the possible shapes of a GADT's container we must
pattern match on the type of the data to be contained. For this, it is
essential that a GADT calculus support an equality type
$\mathsf{Equal}$ that is a singleton set when its two type arguments
are the same and is the empty set otherwise. That is, the type
$\mathsf{Equal}$ must be the syntactic reflection of semantic equality
function $\mathit{Equal}$. The type $\mathsf{Equal}$ can be
defined via GADT syntax as
\[\begin{array}{l}
\mathsf{data\, Equal\,a\,b\,where}\\
\mathsf{\;\;\;\;\;\;\;\;Refl ::\, Equal\,c\,c}\\
\end{array}\]
\noindent
Its Church encoding (in a higher-kinded calculus such as
$F_\omega$~\cite{atk12}) is
\[\mathsf{Equal\,a\,b} = \mathsf{\forall f.\, (\forall c.\,
  f\,c\,c) \to f\,a\,b}\] Following the technique in Section~4.5
of~\cite{atk12}, we can rewrite its Church encoding as
\[\mathsf{Seq\,a} = \mathsf{\forall f.\,(\forall d. \,(f\,d + \exists b
  c.\, f\, b \times f \,c \times \mathsf{Equal}\, d \,(b \times c))
  \to f \,d) \to f\, a}\]
which is logically equivalent to the more intuitive encoding
\[\mathsf{Seq\,a} = \mathsf{\forall f.\, (\forall d.\,d \to f\, d)
  \to (\forall b\,c\,d.\,Equal\,d\,(b\times c) \to f\,b \to f\,c \to
  f\,d) \to f\,a}\]

Importantly, the function $\mathit{Equal}\,a$ cannot be made into a
functor. Equivalently, $\mathsf{Equal\,a}$ cannot be made an instance
of Haskell's $\mathsf{Functor}$ class. Indeed, if $\mathsf{Equal\,a}$
supported a function
\[\mathsf{map_{Equal\,a} :: (b \to c) \to Equal\,a\,b \to
  Equal\,a\,c}\]
then defining
\[\begin{array}{l}
\mathsf{eqElim :: Equal\, a\, b \to b \to a}\\
\mathsf{eqElim\, Refl\, x = x} 
\end{array}\]
would allow us to construct
an element
\[\mathsf{eqElim\, (map_{Equal\,0}\, absurd\, Refl)\, 1}\]
of the empty type $\mathsf{0}$, where $\mathsf{absurd :: 0 \to c}$ is
the
empty function from  $\mathsf{0}$
into
$\mathsf{c}$.
But this is not possible.

Since the Church encoding of a GADT that is not a nested type involves
the equality type, its $\mathsf{map}$ function must necessarily
involve the equality type's $\mathsf{map}$ function. But since the
equality type does not support a $\mathsf{map}$ function,
it is immediate that
a Church encoding of a GADT that is not a nested type cannot
support a $\mathsf{map}$ function either.
The underlying problem is illustrated for the GADT $\mathsf{Seq}$ in
Section~\ref{sec:intro}. In that setting, given a function $\mathsf{f
  :: (a \times b) \to e}$, the term $\mathsf{map_{Equal\,(a \times
    b)}\,f\, Refl}$ would have type $\mathsf{Equal\,(a \times b)\,e}$.
But we have no way to produce a term of this type in the absence of a
functorial $\mathsf{map}$ function for $\mathsf{Equal}$, and thus no
way to produce a term of type $\mathsf{Seq\,e}$ using the
$\mathsf{Pair}$ constructor, as would be required by the clause of
$\mathsf{map_{Seq}}$ for $\mathsf{Pair}$.

\subsection{GADTs as Fixpoints}\label{sec:prim-gadts}

The Church encoding of a GADT corresponds to the data type comprising
just those elements specified by the GADT's syntax. By contrast, the
fixpoint representation of a GADT corresponds to the data type
comprising all data elements in the functorial completion of the
GADT's syntax. In this latter reading, GADTs can, like ADTs, be
modeled as fixpoints of higher-order functors. Syntactically, {\em
  higher-orderness} is essential; since the type arguments to the GADT
being defined are not necessarily uniform across all of its instances
in the types of its data constructors, GADTs cannot be seen as
first-order fixpoints the way ADTs can. Semantically, the use of
(higher-order) {\em functors} is essential, as in the case of ADTs, to
guarantee the existence of the (higher-order) fixpoints being
computed~\cite{tfca}.

To illustrate, consider again the GADT $\mathsf{Seq}$. Because its
type argument varies in the instances of $\mathsf{Seq}$ appearing in
the types of its data constructor $\mathsf{Pair}$, $\mathsf{Seq}$
cannot be modeled as the fixpoint of any first-order functor. As shown
in~\cite{jp19}, it can, however, be modeled as a solution to the
higher-order fixpoint equation
\[H\,f\,b = b \,+\,(Lan_{\lambda c d. c \times d}\, \lambda
c d. f c \times f d)\,b\] where $Lan_K \,F$ is the left Kan extension
of the functor $F$ along the functor $K$. In general, the left Kan
extension $Lan_K \,F : {\cal E} \to {\cal D}$ of $F : {\cal C} \to
{\cal D}$ along $K : {\cal C} \to {\cal E}$ is the best functorial
approximation to $F$ that factors through $K$. Intuitively, ``best functorial
approximation'' means that $Lan_K \,F$ is the smallest functor that
both extends the image of $K$ to $\cal D$ and agrees with $F$ on $\cal
C$, in the sense that, for any other such functor $G$, there is a
morphism of functors (i.e., a natural transformation) from $Lan_K \,F$
to $G$. Formally, this is captured by the following
definition~\cite{mac71}:
\begin{definition}\label{def:lke}
If $F : {\cal C} \to {\cal D}$ and $K : {\cal C} \to {\cal E}$ are
functors, then the {\em left Kan extension of $F$ along $K$} is a
functor $\mathit{Lan}_K\,F : {\cal E} \to {\cal D}$ together with a
natural transformation $\eta : F \to (\mathit{Lan}_K\,F) \circ K$ such
that, for every functor $G : {\cal E} \to {\cal D}$ and natural
transformation $\gamma : F \to G \circ K$, there exists a unique
natural transformation $\delta : \mathit{Lan}_K\,F \to G$ such that
$(\delta K) \circ \eta = \gamma$. This is depicted in the diagram
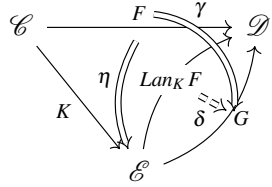
\[\begin{tikzcd}[row sep = huge]
{\cal C}
\ar[rr, "{F}"{name=Fa, above}, ""{name=F, below}]
\ar[rd, "{K}"']
&& {\cal D} \\
& {\cal E}
\ar[Rightarrow, bend right = 25, from=F, "{\eta}"']
\ar[ur, bend left, "{\mathit{Lan}_{K}\,F}"{name=Lan, description}]
\ar[ur, bend right, "{G}"'{name=L, right}, ""'{name=Lr, right}]
\ar[Rightarrow, bend left = 45, from=Fa, to=Lr, "{\gamma}" near start]
\ar[Rightarrow, dashed, from=Lan, to=L, "{\delta}"']
\end{tikzcd}\]
\end{definition}

To represent GADTs as fixpoints in a setting in which types are
interpreted as sets, a calculus must support a primitive construct
$\mathsf{Lan}$ such that the type constructor $\mathsf{Lan_K\,F}$ is
the syntactic reflection of the left Kan extension $\Lan_K \,F$ of the
functor $F$ interpreting $\mathsf{F}$ along the functor $K$
interpreting $\mathsf{K}$. In this setting, the categories
$\mathcal{C}$, $\mathcal{D}$, and $\mathcal{E}$ must all be of the
form $\mathit{Set}^n$ for some $n$. Using $\mathsf{Lan}$ we can then
rewrite the type of a constructor $\mathsf{C :: F\, a \to G\,(K\,a)}$
as $\mathsf{C :: (Lan_K\,F)\,a \to G\,a}$ since morphisms (i.e.,
natural transformations) from $F$ to $G \circ K$ are in one-to-one
correspondence with those from $\mathit{Lan}_K\,F$ to $G$ (see
e.g.,~\cite{rie16}). That is, writing $F \Rightarrow G$ for the set of
natural transformations from a functor $F$ to a functor $G$, we have
\begin{equation}\label{eq:nat-transfs}
F \Rightarrow G \circ K\; \simeq \; Lan_K\,F \Rightarrow G
\end{equation}
The calculus must also support a primitive type
constructor $\mathsf{\mu}$ that is the syntactic reflection of the
(now higher-order) fixpoint operator on
$\mathit{Set}^\mathit{Set}$. Using $\mu$ and $\Lan$ we can then
represent a GADT as a higher-order fixpoint. For example,
we can represent the GADT $\mathsf{Seq}$ as
\[\mathsf{Seq\,a} = \mathsf{(\mu \phi.\lambda b.\, b + (Lan_{\lambda c
    d. c \times d} \lambda c d. \phi c \times \phi d)\,b)\,a}\] The
fact that $Lan_K\,F$ is the best functorial approximation to $F$
factoring through $K$ means that the type constructor
$\mathsf{Lan_K\,F}$ computes the smallest collection of data that is
generated by the corresponding GADT data constructor's syntax and also
supports a $\mathsf{map}$ function. The fixpoint representation of any
GADT thus comprises the smallest data type that both includes the data
specified by that GADT's syntax and also supports a $\mathsf{map}$
function.
When viewed as fixpoints, then,
GADTs are generally underspecified by
their syntax.

\vspace*{0.05in}

In this section we have seen that GADTs can be represented either as
Church encodings or as fixpoints. This exactly mirrors the situation
for ADTs and nested types described in Sections~\ref{sec:adts}
and~\ref{sec:gadts}. But whereas the two representations are always
semantically equivalent for ADTs and nested types, they are not, in
general, semantically equivalent for GADTs. This will be shown
formally in the next section.

\section{Non-Equivalence of Church Encodings and 
    Fixpoint Representations of GADTs}\label{sec:non-equiv}

To see that the Church encoding of a GADT and its fixpoint
representation need not be semantically equivalent, consider again the
GADT $\mathsf{G}$ defined in~\eqref{eq:G}. Despite its simplicity,
this GADT serves as an informative case study highlighting the
differences between a GADT's Church encoding and its fixpoint
representation --- even if we consider only the data elements it
contains, and ignore whether or not it supports a $\mathsf{map}$
function.

\begin{example}\label{ex:g1}

Syntactically, the GADT $\mathsf{G}$ defined in~\eqref{eq:G} comprises
a single data element, namely $\mathsf{C :: G \, 1}$. This is captured
by $\mathsf{G}$'s Church encoding $\mathsf{G\,a} = \mathsf{\forall\,f.
  \, (\forall\,c.\, Equal\,1\,c \to f\,c) \to f\,a}$ from
~\cite{atk12}, which is equivalent in the parametric model given there
to
\[\mathsf{G\, a = Equal\,1\,a}\]
As the above equation makes clear,
$\mathsf{G}$'s
effect is simply to test its argument for equality against the unit
type $\mathsf{1}$. The interpretation of the Church encoding of
$\mathsf{G}$ is thus, as explained in Section~\ref{sec:CEs}, the
function whose value is a singleton set when the interpretation of
$\mathsf{a}$ is $1$ and the empty set otherwise. This perfectly
accords with $\mathsf{G}$'s syntax, which indeed delivers just one
data element when $\mathsf{a}$ is $\mathsf{1}$ and no elements
otherwise.

To compute the interpretation of $\mathsf{G}$'s fixpoint
representation we first note that the type of $\mathsf{G}$'s solitary
constructor $\mathsf{C :: G\,1}$ is equivalently expressed as
$\mathsf{C :: 1 \to G\,1}$ or, using~\eqref{eq:nat-transfs}, as
$\mathsf{C :: (Lan_{\lambda u. 1}\,\lambda u. 1)\,a \to G\, a}$, where
$\mathsf{\lambda u. 1}$ is the syntactic reflection of the constantly
$1$-valued functor from the category $\mathit{Set}^0$ with a single
object to $\mathit{Set}$. We can therefore represent $\mathsf{G}$ as
\[\mathsf{G\,a} = \mathsf{(\mu \phi. \lambda b.
 (Lan_{\lambda u. 1} \,\lambda u.  1)\,b)\,a}\] The interpretation of
$\mathsf{G}$ is therefore obtained by computing the fixpoint of the
interpretation of the body $\mathsf{\lambda b. (Lan_{\lambda u. 1}
  \,\lambda u. 1)\,b}$ of the syntactic fixpoint $\mathsf{\mu
  \phi. \lambda b.  (Lan_{\lambda u. 1} \,\lambda u.  1)\,b}$ and
applying the result to $\mathsf{a}$. But since the recursion variable
$\mathsf{\phi}$ does not appear in this body, the interpretation of
the fixpoint is just the interpretation of the body itself.  The
interpretation of $\mathsf{G\,a}$ is therefore $(\mathit{Lan}_{\lambda
  u. 1} \lambda u. 1)\, A$, where $A$ interprets $\mathsf{a}$.
It turns out, however, that, for any set $A$, $(\mathit{Lan}_{\lambda
  u. 1} \lambda u. 1)\,A$ is, in fact, exactly $A$. Indeed,
Proposition~7.1 of~\cite{blw03} gives that $(\mathit{Lan}_{\lambda
  u. 1} \lambda u. 1)\,A$ can be computed as
\[\big( \bigcup_{U : \mathit{Set}^0,\,f: (\lambda u.\,1)\,U \to A} (\lambda
u. 1)\,U \;\big) \,/\,\sim
\;\;\;\;=\;\;\;\;
\big( \bigcup_{U :
  \mathit{Set}^0,\,f: 1 \to A} 1 \big)\,/\,\sim\]
where $U$ is the unique object of $\mathit{Set}^0$, $*$ is the unique
element of the singleton set $1$, and $\sim$ is the smallest
equivalence relation such that $(U,f, *)$ and $(U,f',*)$ are related
if
\[
\begin{tikzcd}
(\lambda u.1)\,U \ar[rr, "{\mathit{(\lambda u. 1)\,id_U}}"] \ar[dr, "f"']
&& (\lambda u.1)\,U \ar[dl, "f'"] \\
&A
\end{tikzcd} \;\;\; =\;\;\;
\begin{tikzcd}
1 \ar[rr, "{\mathit{id_1}}"] \ar[dr, "f"']
&&1 \ar[dl, "f'"] \\
&A
\end{tikzcd}
\]
commutes, i.e., if $f = f'$. Since the relation generating $\sim$ is
already an equivalence relation, we have that $(U,f,*) \sim (U,f', *)$
iff $f = f'$. Thus, up to isomorphism, $(\mathit{Lan}_{\lambda u. 1}
\lambda u. 1)\,A = \{ f : 1 \to A\}$, i.e., $(\mathit{Lan}_{\lambda
  u. 1} \lambda u. 1)\,A = A$. This is different from the
interpretation of $\mathsf{G}$'s Church encoding from the previous
paragraph.

Putting it all together, we see that the Church encoding of
$\mathsf{G}$ and its fixpoint representation are not semantically
equivalent: the interpretation of the former has exactly one data
element at instance $\mathsf{G\,1}$ and no elements at any other
instances, whereas the interpretation of the latter has data elements
at every instance other than $\mathsf{G\,0}$.
These additional data elements can be obtained by reflecting back into
syntax the elements $\mathit{map}_G\,f_a\,c \in G\,A$ resulting from
applying the functorial action $\mathit{map}_G$ of $\mathsf{G}$'s
interpretation $G$ to the functions $f_a : 1 \to A$ determined by the
elements $a$ of $A \not = \emptyset$ and the interpretation $c$ of
$\mathsf{C}$.
\end{example}

Forced to choose, a programmer would likely find the idea that a GADT
contains data not specified by its syntax more than a little
disturbing. What, they might ask, should a data type contain other
than data that are constructed using its data constructors? Why should
a GADT contain ``hidden'' elements that are not specified by the
GADT's syntax and are only accessible via applications of
$\mathsf{map}$? From a semanticist's point of view, however, the
primitive representation of GADTs is entirely reasonable. Indeed, they
would likely find the nonfunctorial nature of a GADT's Church encoding
unnerving at best. After all, they would likely argue, the data in a
GADT shouldn't change or become ill-typed just because a function is
mapped over it. The fact that this happens when GADTs are represented
by their Church encodings actually highlights how GADTs {\em do not}
generalize the essential, container-ish nature of ADTs at all. A
semanticist might therefore conclude that GADTs are seriously
misnamed.

Because they are not semantically equivalent, the two representations
of GADTs have very different implications for parametricity. We
explore the differences in parametricity results for the two
representations of GADTs in the next section.

\section{Parametricity in the Presence of GADTs}\label{sec:par}
{\em Relational parametricity} encodes a powerful notion of type
uniformity, or representation independence, for data types in
functional languages. It formalizes the intuition that a polymorphic
program must act uniformly on all of its possible type instantiations
by requiring that every such program preserves all relations between
pairs of types at which it is instantiated. Parametricity was
originally put forth by Reynolds~\cite{rey83} for System F. It was
later popularized as Wadler's ``theorems for free''~\cite{wad89}, so
called because it can deduce properties of programs solely from their
types, i.e., with no knowledge whatsoever of the text of the programs
involved.  Most of Wadler's free theorems are consequences of
naturality for polymorphic list-processing functions. However,
parametricity can also derive results that go beyond just naturality,
such as inhabitation results, and prove the equivalence of Church
encodings and fixpoint representations of ADTs and
nested types by validating short cut fusion and other program
equivalences for them.

To discuss relational parametricity we will need to interpret data
types not just in $\mathit{Set}$, but in a suitable category of
relations as well. The following definition is standard:
\begin{definition}\label{def:rel}
  The category $\mathit{Rel}$ has:
  \begin{itemize}
\item objects: A relation is a triple $(A,B,R)$, where $R$ is a
  subset of $A \times B$.
\item morphisms: A morphism from $(A,B,R)$ to $(A',B',R')$ is a pair
  $(f : A \to A',g : B \to B')$ of functions in $\mathit{Set}$ such
  that $(f a,g\,b) \in R'$ if $(a,b) \in R$.
\item identities: The identity morphism on $(A,B,R)$ is the pair
  $(\mathit{id}_A : A \to A, \mathit{id}_B : B \to B)$.
\item composition: Composition is the componentwise composition in
  $\mathit{Set}$. That is, $(g_1,g_2) \circ (f_1,f_2) = (g_1 \circ
  f_1, g_2 \circ f_2)$, where the composition being defined on the
  left-hand side is in $\mathit{Rel}$, and the two componentwise
  compositions on the right-hand side are in $\mathit{Set}$.
\end{itemize}
\end{definition}
\noindent
We write $R : \mathit{Rel}\,(A,B)$ for $(A,B,R) \in \mathit{Rel}$.  If
$R : \mathit{Rel}\,(A,B)$ then we write $\pi_1 R$ and $\pi_2 R$ for
the {\em domain} $A$ and {\em codomain} $B$ of $R$, respectively. We
write $\mathit{Eq}_A = (A,A,\{(x,x)~|~ x \in A\})$ for the {\em
  equality relation} on the set $A$.

The key idea underlying parametricity is to give each type
$\mathsf{G[a]}$\footnote{The notation $\mathsf{G[a]}$ indicates that
  $\mathsf{G}$ is a type with one hole which has been filled with the
  type $\mathsf{a}$.}  with one free variable $\mathsf{a}$ a {\em set
  interpretation} $G_0$ taking sets to sets and a \emph{relational
  interpretation} $G_1$ taking relations $R : \mathit{Rel}\,(A,B)$ to
relations $G_1 \,R : \mathit{Rel}\,(G_0 \,A, G_0 \,B)$, and to
interpret each term $\mathsf{t\,(a,x) :: G[a]}$ with one free term
variable $\mathsf{x :: F[a]}$ as a function $t$ associating to each
set $A$ a morphism $t \,A : F_0\,A \to G_0\,A$ in
$\mathit{Set}$. Here, $F_0$ is the set interpretation of $\mathsf{F}$.
These interpretations are given inductively on the structures of
$\mathsf{G}$ and $\mathsf{t}$ in such a way that they imply two
fundamental theorems. The first is an \emph{Identity Extension Lemma},
which states that $G_1\,\mathit{Eq}_A = \mathit{Eq}_{G_0 A}$, and is
the essential property that makes a model relationally parametric
rather than just induced by a logical relation.  The second is an
\emph{Abstraction Theorem}, which states that, for any $R
:\mathit{Rel}\,(A, B)$, $(t\, A, t\,B)$ is a morphism in
$\mathit{Rel}$ from $(F_0\,A,F_0\,B,F_1\,R)$ to
$(G_0\,A,G_0\,B,G_1\,R)$. The Identity Extension Lemma is similar to
the Abstraction Theorem except that it holds for {\em all} elements of
a type's interpretation, not just those that interpret terms.  Similar
theorems are required for types and terms with any number of free
variables. In particular, if $\mathsf{t}$ is closed (i.e., has no free
term variables) then $t\,A \in G_0\,A$.

The existence of parametric models that interpret GADTs as the
interpretations of their Church encodings follows from, e.g., the
existence of the parametric model of $F_\omega$ constructed
in~\cite{atk12}.  In that model, types are interpreted ``in parallel''
in (types corresponding to) $\mathit{Set}$ and $\mathit{Rel}$ in the
usual way, including the familiar ``cutting down'' of the
interpretations of $\forall$-types to just those elements that are
``parametric''~\cite{rey83,wad89} to ensure that the Identity
Extension Lemma holds. If the set interpretation of $\mathsf{Equal}$
is the function $\mathit{Equal}$ and if the relational interpretation
of a closed type $\mathsf{a}$ with interpretation $A$ is
$\mathit{Eq_A}$ as intended, then the parametricity property for a
GADT is an inhabitation result saying that the set interpreting any
instance of that GADT contains exactly the interpretations of the data
elements that can be formed using its data constructors and whose type
is that instance. The parametricity property for the Church encoding
of a GADT $\mathsf{G}$ gives that $\mathsf{G\,a}$ is inhabited iff data
elements of the instance of $\mathsf{G}$ at $\mathsf{a}$ can be formed
using $\mathsf{G}$'s data constructors.  In particular, the
parametricity property for the GADT $\mathsf{G}$ from~\eqref{eq:G}
gives that $\mathsf{G\,a}$ contains a single data element if
$\mathsf{a}$ is semantically equivalent to $\mathsf{1}$ and none
otherwise:

\begin{example}\label{ex:CE-par}
Let $\mathsf{t}$ be a closed term of type $\mathsf{G\,a}$ for the GADT
$\mathsf{G}$ defined in~\eqref{eq:G}, let $G = (G_0,G_1)$ be the
interpretation of the Church encoding of $\mathsf{G}$, let $t$ be the
interpretation of $\mathsf{t}$, and let $R :
\mathit{Rel}\,(A,B)$. Then $t\, A \in G_0\,A$ and $t\,B \in G_0\,B$
and, by the Abstraction Theorem (Theorem~3) in~\cite{atk12}, $t\,A$
and $t\,B$ must be related in $G_1\,R$. However, under the semantics
given in~\cite{atk12}, which includes the aforementioned
interpretations of $\mathsf{Equal}$ and closed types, the relational
interpretation $G_1\,R$ of $\mathsf{G}$ is itself $\mathit{Eq}_1$ when
$R$ is the relational interpretation $\mathit{Eq}_1$ of $\mathsf{1}$,
and the empty relation whenever $R$ differs from
$\mathit{Eq}_1$. Reflecting back into syntax we deduce that there can
be no term in the type that is the Church encoding of $\mathsf{G\,a}$
unless $\mathsf{a}$ is semantically equivalent to $\mathsf{1}$.
\end{example}

For fixpoint representations of GADTs the story is completely
different. If, as intended, the set interpretation of
$\mathsf{Lan_K\,F}$ is $\mathit{Lan}_K\,F$, where $K$ is the set
interpretation of $\mathsf{K}$ and $F$ is the set interpretation of
$\mathsf{F}$, then the exact same reasoning gives that the relational
interpretation of $\mathsf{Lan_K\,F}$ is $\mathit{Lan}_K\,F$, where
$K$ is the relational interpretation of $\mathsf{K}$ and $F$ is the
relational interpretation of $\mathsf{F}$. But under these
interpretations there can be no parametric model. The following
counterexample establishes this surprising result. It is the main
technical contribution of this paper.

\begin{example}\label{ex:prim-par}
In any parametric model we must give both a set interpretation and a
relational interpretation for every type as described at the start of
this section. In particular, for every GADT $\mathsf{G}$ we must give
an interpretation $G = (G_0,G_1)$ such that, for every relation $R :
\mathit{Rel}\,(A, B)$, we have $G_1\,R : \mathit{Rel}\,(G_0\,A,
G_0\,B)$.

Intuitively, when $\mathsf{G}$ is viewed as a fixpoint, its data
elements include those given by functorial completion. Since $G_1$ is
a functor, given any relation $S : \mathit{Rel}\,(C, D)$ and any
morphism $m : S \to R$, $G_1\, R$ must contain all elements of the
form $G_1\,m\,x$ for $x \in G_1\,S$.  But the two components $m_1 : C
\to A$ and $m_2 : D \to B$ of $m$ cannot be given independently of one
another, since Definition~\ref{def:rel} entails that
$(m_1\,c, m_2\,d)$ must be in $R$ whenever $(c,d)$ is in $S$ for
$(m_1,m_2)$ to be a well-defined morphism of relations.  The domain of
$G_1\, R$ thus depends on both $A$ and $B$, rather than simply on $A$.
Likewise, the codomain of $G_1\,R$ also depends on both $A$ and $B$.
The domain and codomain therefore cannot simply be $G_0\, A$ and
$G_0\, B$, respectively.  This suggests that GADTs might fail to have
relational interpretation, and thus might fail to have a parametric
model, as described in the previous paragraph.

We can make this informal argument formal by providing a concrete
counterexample.  Consider again the GADT $\mathsf{G}$ given
by~\eqref{eq:G}. The set interpretation of $\mathsf{G}$ is
$\mathit{Lan}_{\lambda u\!. 1}\,\lambda u\!. 1$, i.e., is, by the
reasoning of Example~\ref{ex:g1}, the identity functor on
$\mathit{Set}$. By the exact same reasoning, this time in
$\mathit{Rel}$ rather than in $\mathit{Set}$, the relational
interpretation of $\mathsf{G}$ is $\mathit{Lan}_{\lambda
  u\!.\,\mathit{Eq}_1} \,\lambda u\!.\,\mathit{Eq}_1$, where $\lambda
u\!.\,\mathit{Eq}_1$ is the constantly $\mathit{Eq}_1$-valued functor
from the category $\mathit{Rel}\,^0$ with a single object to
$\mathit{Rel}$.  Indeed, the interpretation is still a left Kan
extension, but now it is the left Kan extension determined by the
functor interpreting the type $\mathsf{\lambda u\!. 1}$ in
$\mathit{Rel}$. For the Identity Extension Lemma to hold we would need
that, for every relation $R : \mathit{Rel}\,(A, B)$, we have that
$(\mathit{\Lan}_{\lambda u\!.\,\mathit{Eq}_1} \,\lambda
u\!.\,\mathit{Eq}_1)\, R$ is a relation between the sets
$(\mathit{Lan}_{\lambda u. 1}\,\lambda u\!. 1)\, A$ and
$(\mathit{Lan}_{\lambda u. 1}\,\lambda u\!. 1)\,B$, i.e., between the
sets $A$ and $B$. However,
this need not be the case.

Consider the relation $R = (1, 2, 1 \times 2)$, where $1 \times 2$
relates the single element of $1$ to both elements of the two-element
set $2$. We expect $(\mathit{Lan}_{\lambda u\!.\,\mathit{Eq}_1}
\lambda u\!.\,\mathit{Eq}_1)\, R$ to be a relation with domain
$1$. Since left Kan extensions preserve projections~\cite{rie16}, we
can compute the domain as
\[\pi_1\,\big( (\mathit{Lan}_{\lambda u\!.\,\mathit{Eq}_1} \lambda
u\!.\,\mathit{Eq}_1) \, R \big)\;=\; (\mathit{Lan}_{\lambda
  u\!.\,\mathit{Eq}_1} \lambda u\!. \,1)\, R\] (Note that the left Kan
extension of $\lambda u\!. \,1$ along $\lambda u\!.\,\mathit{Eq}_1$ is
a functor from $\mathit{Rel}$ to $\mathit{Set}$.)  By the same
reasoning as in Example~\ref{ex:g1}, Proposition~7.1 of~\cite{blw03}
gives that $(\mathit{Lan}_{\lambda u. \mathit{Eq}_1} \lambda u. 1)\,R$
can be computed as
\[\big( \bigcup_{U : \mathit{Rel}^0,\,m: (\lambda u.\,\mathit{Eq}_1)\,U \to R} (\lambda
u. 1)\,U \;\big) \,/\,\approx \;\;\;\;=\;\;\;\; \big( \bigcup_{U :
  \mathit{Rel}^0,\,m: \mathit{Eq}_1 \to R} 1 \big)\,/\,\approx\] where
$U$ is the unique object of $\mathit{Rel}^0$, $*$ is the unique
element of the singleton set $1$, and $\approx$ is the smallest
equivalence relation such that $(U,m,*)$ and $(U,m',*)$ are related if
\[
\begin{tikzcd}
(\lambda u.\mathit{Eq}_1)\,U \ar[rr, "{\mathit{(\lambda
        u.\mathit{Eq}_1)\,id_U}}"] \ar[dr, "m"'] 
&& (\lambda u. \mathit{Eq}_1)\,U \ar[dl, "m'"] \\
&R
\end{tikzcd} \;\;\; =\;\;\;
\begin{tikzcd}
\mathit{Eq}_1 \ar[rr, "{\mathit{id_{\mathit{Eq}_1}}}"] \ar[dr, "m"']
&&\mathit{Eq}_1 \ar[dl, "m'"] \\
&R
\end{tikzcd}
\]
commutes, i.e., if $m = m'$. Since the relation generating $\approx$
is already an equivalence relation, we have that $(U,m,*) \approx
(U,m', *)$ iff $m = m'$. Thus, up to isomorphism,
$(\mathit{Lan}_{\lambda u\!.\,\mathit{Eq}_1} \lambda u\!. \,1)\, R =
\{ m : \mathit{Eq}_1 \to R\}$.  But this set is $\{(!, k_0), (!,
k_1)\}$, where $k_0, k_1 : 1 \to 2$ are the constantly $0$-valued and
$1$-valued functions in $\mathit{Set}$, respectively, and therefore
$\pi_1\,\big( (\mathit{Lan}_{\lambda u\!.\,\mathit{Eq}_1} \lambda
u\!.\,\mathit{Eq}_1) \, R \big)$ is not $1$, as would be needed for
the Identity Extension Lemma to hold.  Since the Identity Extension
Lemma does not hold for models in which GADTs are interpreted by the
interpretations of their fixpoint representations, such models cannot
possibly be parametric.
\end{example}

\vspace*{-0.05in}

We note that it is actually possible to construct a simpler
counterexample to the Identity Extension Lemma for fixpoint
representations of GADTs using the relation $R = (1,\emptyset,
\emptyset)$. However, this relation is somewhat artificial, in the
sense that its domain is larger than is strictly necessary to define
an empty relation. We therefore give the above example using the
relation $(1,2,1 \times 2)$ instead.

Taken together, Examples~\ref{ex:CE-par} and~\ref{ex:prim-par} show
that Church encodings and fixpoint representations of GADTs behave
very differently with respect to parametricity.  This contrasts
sharply with the fact that both ADTs and nested types have the same
parametricity properties regardless of whether they are represented as
their Church encodings or their fixpoint representations.

\section{Conclusion and Related Work}

We have shown that GADTs can be considered as data types in two
contrasting ways: as their Church encodings, in which case they are
completely determined by their syntax, or as fixpoints, in which case
they are regarded as the functorial completion of their syntax. But
regardless of which representation we choose, GADTs fail to have {\em
  all} expected parametricity properties.  Indeed, we can read the
results of Section~\ref{sec:par} as showing that {\em i}) inhabitation
and other non-naturality results for calculi representing GADTs as
Church encodings cannot be extended to calculi representing them as
fixpoints, and {\em ii}) naturality results for calculi representing
GADTs as fixpoints cannot be extended to calculi representing them as
Church encodings. So, neither representation guarantees {\em both}
naturality consequences {\em and} non-naturality consequences of
parametricity. Intuitively, this is because a GADT is not a data type
in the usual container-ish sense, and completing it so that it becomes
one destroys its relationally uniform --- i.e., parametric ---
behavior.  Which representation of data types a user chooses can
therefore have significant consequences for programming with, and
reasoning about, GADTs.

Although there are no parametric models for calculi representing GADTs
as fixpoints, GADT analogues of most of Wadler's free theorems for
ADTs still hold.
Indeed, most of the parametricity results in Figure~1 of~\cite{wad89}
are actually naturality properties.
Naturality properties are often regarded as consequences of {\em
  parametricity}, but are, in fact, derivable directly from data
types' {\em functorial semantics}. Even for data types whose Church
encodings and fixpoint representations coincide, the analysis in this
paper clearly distinguishes those free theorems that are actually
consequences of naturality from those that are true consequences of
parametricity.

There are treatments of GADTs beyond those discussed in the main body
of this paper.  Atkey's parametric model for $F_\omega$
from~\cite{atk12} represents data types --- including GADTs --- as
Church encodings. It requires the user to supply a $\mathsf{map}$
function for the type constructor whose fixpoint characterizes the
data type. But, importantly, functoriality of an underlying type
constructor does not imply functoriality of its fixpoint, so the data
type itself still need not necessarily support a $\mathsf{map}$ in
Atkey's model. Similarly,~\cite{vw10} presents a parametric model for
an extension of $F_\omega$ that supports type equality, and thus can
encode GADTs, but this model still does not guarantee functoriality;
accordingly, the parametric properties of GADTs described in the
precursor work~\cite{vw06} to~\cite{vw10} are all inhabitation results
rather than naturality results. In~\cite{ms09} GADTs are represented
as Scott encodings rather than Church encodings but, again, only
inhabitation results are cited for them. GADTs are treated explicitly
as fixpoints of discrete functors in~\cite{jg08}, as initial algebras
of dependent polynomial functors in~\cite{gh03,hf11}, and as indexed
containers in~\cite{ma09}. The latter two treatments move toward
seeing GADTs as data types in a dependent type theory. A categorical
parametric model of dependent types has been given in~\cite{agj14},
but, as with the models mentioned above, this model also does not
guarantee that GADTs have functorial semantics.

\vspace*{0.2in}

\noindent {\bf Acknowledgment} This work was supported by National
Science Foundation award 1713389.

\bibliographystyle{eptcs}
\bibliography{gadt-story-references}

\end{document}

%% file: macros.tex
\DeclareMathAlphabet{\mathpzc}{OT1}{pzc}{m}{it}

\newcommand{\id}{id}

\newcommand{\nat}{\mathpzc{Nat}}

\newcommand{\List}{\mathsf{List}\,}

\newcommand{\Lan}{\mathsf{Lan}}

\renewcommand{\nat}{\mathbb{N}}

\newcommand{\ar}[1]{\##1}

\newcommand{\greyout}[1]{{\color{gray} {#1}}}

%% file: mytheorem.tex
\newtheorem{thm}{Theorem}
